\documentclass[%
article,
twocolumn,
amsmath,amssymb,
aps,
journal=prl,
]{revtex4-2}

\usepackage{natbib}
\usepackage{graphicx}
\usepackage{dcolumn}
\usepackage{bm}
\usepackage{xcolor}
\usepackage{float}
\usepackage{newtxtext,newtxmath}

\begin{document}

\title{
Nonperturbative Simulation of Anharmonic Rattler Dynamics in Type-I Clathrates with Vibrational Dynamical Mean-Field Theory
}

\author{Dipti Jasrasaria}
\email{dj2667@columbia.edu}
\address{Department of Chemistry, Columbia University, New York, New York 10027, USA}

\author{Timothy C. Berkelbach}
\email{t.berkelbach@columbia.edu}
\affiliation{Department of Chemistry, Columbia University, New York, New York 10027, USA}

\date{\today}

\begin{abstract}
We use vibrational dynamical mean-field theory (VDMFT) to study the vibrational structure of type-I clathrate solids, specifically X$_8$Ga$_{16}$Ge$_{30}$, where X=Ba,Sr. These materials are cage-like chemical structures hosting loosely bound guest atoms, resulting in strong anharmonicity, short phonon lifetimes, and ultra-low thermal conductivities. Presenting the methodological developments necessary for this first application to three-dimensional, atomistic materials, we validate our approach through comparison to molecular dynamics simulations and show that VDMFT is extremely accurate at a fraction of the cost. 
Through the use of nonperturbative methods, we find that anharmonicity is dominated by four-phonon and higher-order scattering processes, and it causes rattler modes to shift up in frequency by 50\% (10~cm$^{-1}$) and to have lifetimes of less than 1~ps; this behavior is not captured by traditional perturbation theory. Furthermore, we analyze the phonon self-energy and find that anharmonicity mixes guest rattling modes and cage acoustic modes, significantly changing the character of the harmonic phonons.
\end{abstract}

\maketitle


Anharmonicity in the lattice vibrations of solids is responsible for temperature-dependent phonon frequency shifts and lifetimes, thermal expansion, and crystal structure stability~\cite{Cowley1963}. A microscopic understanding of anharmonicity is essential for the emerging field of phononics, which aims to design and control the structural and dynamical properties of materials~\cite{li2011phononics, maldovan2013sound}. Examples include the engineering of materials with low thermal conductivities~\cite{qian2021phonon}, which are important for thermal insulation or the generation of electricity from waste heat via thermoelectrics, or ultrafast optical control of lattice structure and dynamics~\cite{forst2011nonlinear}. From a theoretical perspective, the accurate description of anharmonicity requires the solution of a many-body problem, demanding the development of approximate numerical methods.

The simplest methods for describing anharmonicity are static mean-field theories, such as self-consistent phonon theory~\cite{Hooton1958, KoehlerPRL1966, werthamer1970self, klein1972rise,tadano2018first}, which describe anharmonic systems using effective harmonic Hamiltonians with temperature-dependent frequencies. While these methods successfully predict some thermodynamic properties~\cite{Rudin_prl2008, Mauri_prb2014, TadanoSrTiO3_prb2015, Troster_prb2021, Tadano_prl2022}, they are not able to account for phonon lifetimes or non-quasiparticle (QP) effects. Perturbative methods can be used to calculate lifetimes due to phonon-phonon interactions~\cite{Klemens1966, Walsh_prb2016, Marianetti_prb2023, Amon_prb2009, Allen_prb2010}, but they are often limited to lowest-order perturbation theory (PT) of three-phonon scattering processes and fail for systems with strong anharmonicity. Molecular dynamics (MD) simulations can describe anharmonic effects of classical nuclei exactly~\cite{ladd1986lattice, Amon_prb2009, Allen_prb2010, Simak_prb2011, Simak_prb2011_TDEP, Marianetti_prb2023}, but the computational cost associated with such direct simulation makes them expensive, especially for the large system sizes necessary to eliminate finite-size effects; moreover, nuclear quantum dynamics can only be treated approximately~\cite{markland2018nuclear}.

In this work, we apply the recently developed vibrational dynamical mean-field theory (VDMFT)~\cite{shih_anharmonic_2022}, which is an extension of the successful DMFT for strongly correlated electrons~\cite{GeorgesPhysRevB1992, GeorgesRevModPhys1996, kotliar2004strongly, KotliarRevModPhys2006}. VDMFT provides a nonperturbative description of local anharmonicity, and, as a Green's function theory, naturally yields both phonon frequency shifts and lifetimes. Here, we advance VDMFT by developing the methods necessary for application to three-dimensional atomistic solids with complex unit cells. We apply this method to study the anharmonic vibrational structure of clathrate solids, which are frameworks of covalently bonded atoms that host loosely bound ``guest" atoms within their cage-like structures. The cage-guest interactions are strongly anharmonic, but their spatial locality makes these materials an ideal testbed for VDMFT.

We focus on the type-I clathrates $\text{Ba}_{8}\text{Ga}_{16}\text{Ge}_{30}$ (BaGG) and $\text{Sr}_{8}\text{Ga}_{16}\text{Ge}_{30}$ (SrGG), which have garnered much interest due to their ultra-low thermal conductivities and promise for thermoelectric applications~\cite{nolas_semiconducting_1998, christensen_avoided_2008, dong_chemical_2000, takasu_dynamical_2006, lee_neutron_2007, takabatake_phonon-glass_2014, tadano_impact_2015, tadano_quartic_2018, ikeda_kondo-like_2019, lindroth_thermal_2019, godse_anharmonic_2022}, as well as the fictitious empty clathrate $\text{Ga}_{16}\text{Ge}_{30}$ (GG). Theoretical and experimental studies of BaGG, SrGG, and related materials have revealed hybridization between acoustic modes of the cage lattice and optical, rattling modes of guest atoms, showing an avoided crossing in the harmonic dispersion relation~\cite{christensen_avoided_2008, tadano_impact_2015, tadano_quartic_2018} with potential implications for the thermal conductivity. 
While anharmonicity in these materials has been studied theoretically using analytical models~\cite{baggioli_theory_2019, pailhes_phonons_2023}, MD simulations~\cite{dong_theoretical_2001, xi_off-center_2018}, mean-field theory~\cite{tadano_quartic_2018, godse_anharmonic_2022}, and lowest-order PT~\cite{tadano_impact_2015, tadano_quartic_2018, ikeda_kondo-like_2019,lindroth_thermal_2019, godse_anharmonic_2022}, this work systematically examines anharmonicity with methods that go beyond conventional PT and/or static mean-field theory to determine the significance of nonperturbative effects.

We use VDMFT to calculate the anharmonic spectral functions of GG, BaGG, and SrGG at 300~K, and we find excellent agreement with those calculated from MD simulations at a fraction of the cost. Moreover, we find that lowest-order PT of three-phonon scattering processes fails to describe the short phonon lifetimes predicted by MD and VDMFT for the systems studied in this work. These comparisons validate VDMFT as an efficient and accurate method for describing anharmonicity in real materials beyond PT. While the vibrational structure of empty clathrates is relatively harmonic, our results show that anharmonicity significantly affects lattice dynamics of the filled clathrates studied here, especially SrGG. Guest-dominant phonon modes in particular show large frequency shifts, short lifetimes, and substantial mixing with cage-acoustic modes. 



\begin{figure}[t]
\includegraphics[width=3.375in]{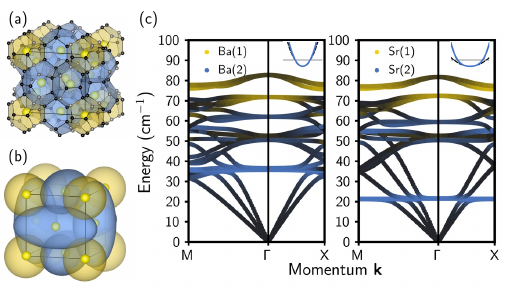}
\caption{(a) Crystal structure of BaGG \cite{bentien_crystal_2005}, where Ba(1) atoms are in dodecahedral cages (gold) and Ba(2) atoms are in tetrakaidecahedral cages (blue). (b) Schematic of coarse-grained model of filled clathrates BaGG and SrGG, where the large cage atoms are colored according to their quartic cage-guest potentials. (c) Harmonic dispersion relations of BaGG (left) and SrGG (right), colored by their atomic character. The insets show the soft anharmonic interaction (blue line) between the X(2) guest atoms and the 24-atom cages as well as the harmonic fitting (black dashed line) and value of $k_BT$ at 300\,K (grey line).}
\label{fig:model}
\end{figure}

To study the vibrational structure of type-I clathrates, we first develop a coarse-grained model of the material, which is shown in Fig.~\ref{fig:model}. Because we are not interested in the high frequency intra-cage dynamics nor the precise locations of the alloyed Ga and Ge atoms, we replace the clathrate cages by single ``hollow'' atoms. For simplicity, these cage atoms interact with one another through identical Lennard-Jones potentials. To ensure the dynamical stability of the crystal, these unified cage atoms are arranged on an FCC lattice, which roughly approximates the positions of the atomistic clathrate cages. At each FCC site, guests are described by smaller atoms that interact with the cage atoms at those sites through anharmonic, quartic potentials. X(1) guest atoms (where X=Ba,Sr) at the cube vertices interact via harder, isotropic potentials, representing interactions with the dodecahedral cages, while X(2) guest atoms at the cube faces interact via softer, anisotropic potentials, mimicking the rattling motions of guests in the tetrakaidecahedral cages \cite{takabatake_phonon-glass_2014}. Thus, a single unit cell in our model consists of eight atoms: four cage atoms and four guest atoms. The complete Hamiltonian for our clathrate model is given by
%
\begin{gather}
\begin{split}
H &= \frac{1}{2}\sum_{\bm{m}\alpha}\left(\frac{\bm{p}_{\bm{m}\alpha}^{2}}{m_{\alpha}}\right)+\frac{1}{2}\sum_{\bm{m}\alpha,\bm{n}\beta}^{\text{cage}} V_{\text{LJ}}\big(\left|\bm{r}_{\bm{m}\alpha}-\bm{r}_{\bm{n}\beta}\big|\right) \\
&\hspace{1em} +\sum_{\bm{m}}\sum_{\alpha}^{\text{cage}}\sum_{\beta}^{\text{guest}}{}^{'} V_{\text{q}}^{\alpha\beta}\left(\bm{r}_{\bm{m}\alpha}-\bm{r}_{\bm{m}\beta}\right)\label{eq:Hamiltonian}
\end{split}\\
V_{\text{LJ}}\left(r\right) = 4\epsilon\left[\left(\frac{\sigma}{r}\right)^{12}-\left(\frac{\sigma}{r}\right)^{6}\right]\label{eq:V_LJ}\\
V_{\text{q}}^{\alpha\beta}\left(\bm{r}\right) = \sum_{i}\left(\frac{1}{2}K_{\beta,i}r_{i}^{2}+g_{\beta,i}r_{i}^{4}\right)\,,\label{eq:V_q}
\end{gather}
where the primed summation indicates that only cages and guests on the same lattice site interact. Here, $\bm{m},\bm{n}$ are lattice translation vectors, $\alpha,\beta$ are indices over atoms in the unit cell, and $i$ is an index over the Cartesian directions. The position, momentum, and mass of atom $\alpha$ in cell $\bm{m}$ are given by $\bm{r}_{\bm{m}\alpha}$, $\bm{p}_{\bm{m}\alpha}$, and $m_{\alpha}$, respectively.

To parameterize the above Hamiltonian, we fit the LJ parameters and the harmonic frequencies, $K_{\alpha,i}$, of the cage-guest interactions to reproduce the harmonic dispersion relations from \textit{ab initio} calculations \cite{tadano_impact_2015, tadano_quartic_2018, godse_anharmonic_2022}.
The harmonic dispersion relation is obtained through diagonalization of the dynamical matrix,
\begin{equation}
\bm{\mathcal{D}}_{\alpha i,\beta j}\left(\bm{k}\right)=\frac{1}{\sqrt{m_{\alpha}m_{\beta}}}\sum_{\bm{m}}e^{i\bm{k}\cdot\left(\bm{R}_{\bm{m}\alpha}-\bm{R}_{\bm{0}\beta}\right)}\frac{\partial^{2}\mathcal{V}}{\partial u_{\bm{m}\alpha i}\partial u_{\bm{0}\beta j}}\,,
\end{equation}
where $\bm{k}$ is a wavevector in the first Brillouin zone (BZ), and $\bm{R}_{\bm{m}\alpha}$ is the equilibrium position of atom $\alpha$ in cell $\bm{m}$. The derivative of the lattice potential, $\mathcal{V}$, with respect to atomic displacements, $u_{\bm{m}\alpha i}=r_{\bm{m}\alpha i}-R_{\bm{m}\alpha i}$, is evaluated at the equilibrium lattice configuration. The dynamical matrix defines the harmonic phonon modes,
\begin{equation}
    u_\lambda(\bm{k}) = N^{-1/2}\sum_{\bm{m}\alpha i} c_{\alpha i,\lambda} (\bm{k}) e^{-i\bm{k}\cdot\bm{R}_{\bm{m}\alpha}}\sqrt{m_\alpha}u_{\bm{m}\alpha i}\,,
\end{equation}
%
where $\bm{c}(\bm{k})$ are the eigenvectors of $\bm{\mathcal{D}}(\bm{k})$.

The harmonic dispersion relations of our model BaGG and SrGG are illustrated in Fig.~\ref{fig:model}c. 
The low-frequency rattling motions of X(2) guest atoms lead to flat modes that cut through the acoustic branches of the cage lattice, leading to the hallmark avoided crossing of these materials. Vibrations of the X(1) guest atoms are higher in frequency, hybridizing with the optical modes of the cage lattice. We complete the parameterization of our Hamiltonian by choosing the quartic anharmonicity parameters, $g_{\alpha,i}$, in Eq. (\ref{eq:V_q}) 
to reproduce the behavior of \textit{ab initio} cage-guest potential energy surfaces for guest atoms Ba and Sr \cite{dong_chemical_2000, godse_anharmonic_2022}. Complete details of our clathrate model and further discussion are given in the Supplemental Material (SM).


To compute the anharmonic lattice dynamics of the clathrate model defined above, we use vibrational dynamical mean-field theory (VDMFT). In VDMFT, we compute the anharmonic phonon Green's function (GF)~\cite{Cowley1963,Mahan2000} of the periodic lattice, $\bm{D}(\bm{k},\omega)$, which encodes phonon frequencies and lifetimes. The spectral part of the GF is experimentally measurable by inelastic neutron scattering, and computationally, the GF can be used to calculate one-body averages and approximations to thermal conductivities. 
Although VDMFT can treat quantum or classical nuclei~\cite{shih_anharmonic_2022}, here we use classical dynamics, as nuclear quantum effects are unimportant for these clathrates at room temperature. Thus, the classical GF is
%
\begin{equation}
    \bm{D}(\bm{k},t) = \frac{\theta(t)}{k_BT} \langle \dot{\bm{u}}(\bm{k},t) \bm{u}^T(-\bm{k},0) \rangle\,,
\end{equation}
where $\bm{D}$ is a matrix, $\bm{u}$ is a column vector, and $\langle \cdot \rangle$ indicates an equilibrium average at temperature $T$.
The Fourier transform of the GF satisfies a Dyson equation,
\begin{equation}
\bm{D}^{-1}\left(\bm{k},\omega\right)=\bm{D}_{0}^{-1}\left(\bm{k},\omega\right)-2\bm{\Omega}\left(\bm{k}\right)\bm{\pi}\left(\bm{k},\omega\right)\,,\label{eq:latticeGF}
\end{equation}
where $\bm{D}_{0}\left(\bm{k},\omega\right)=[\omega^{2}\bm{1}-\bm{\Omega}^{2}\left(\bm{k}\right)]^{-1}$ is the GF of the harmonic lattice, $\bm{\Omega}^{2}\left(\bm{k}\right)$ is the dynamical matrix (so that $\bm{\Omega}^{2}\left(\bm{k}\right)=\bm{\mathcal{D}}\left(\bm{k}\right)$ in the atomic basis), and $\bm{\pi}\left(\bm{k},\omega\right)$ is the self-energy that describes anharmonicity in the lattice. 
VDMFT makes the approximation of a local self-energy, $\bm{\pi}\left(\bm{k},\omega\right)\approx\bm{\pi}\left(\omega\right)$, which we calculate nonperturbatively by solving a so-called impurity problem.
Specifically, VDMFT maps the problem of a periodic lattice onto that of a single unit cell (the ``system'') interacting with a fictitious bath of harmonic oscillators characterized by a tailored spectral density \cite{shih_anharmonic_2022,GeorgesPhysRevB1992,GeorgesRevModPhys1996}. This impurity problem is generally easier to solve than the full periodic problem because of the small number of degrees of freedom in the finite system. 

\begin{figure*}[t]
\includegraphics[width=6.75in]{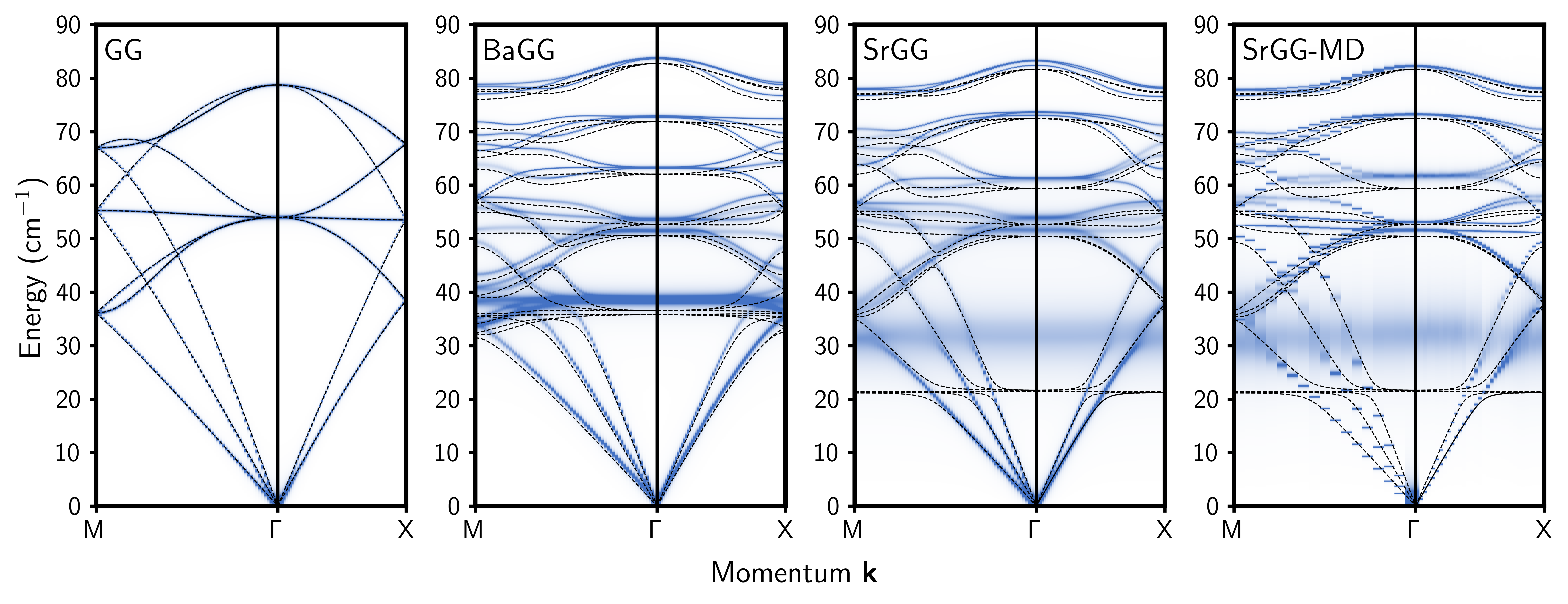}
\caption{Spectral functions of empty and filled clathrates at 300~K calculated using VDMFT. The rightmost panel shows the spectral function of SrGG calculated using MD with a supercell of 8,192 atoms. Black dashed lines indicate the harmonic dispersion relation.}
\label{fig:spectralFunction}
\end{figure*}

The unit cell atoms that constitute the system experience a local, anharmonic, many-body potential, $V_{\text{loc}}(\bm{u})$, and 
the GF that describes the isolated system is given by $\bm{D}_{\text{sys}}\left(\omega\right)=[\omega^{2}\bm{1}-\bm{\Omega}^{2}-2\bm{\Omega}\bm{\pi}\left(\omega\right)]^{-1}$, where $\bm{\Omega}^{2}$ is the system dynamical matrix with respect to $V_{\text{loc}}(\bm{u})$.
The harmonic bath and system-bath coupling are determined by the hybridization, $\bm{\Delta}\left(\omega\right)$, which describes the effect of the lattice on the dynamics of the isolated system,
\begin{equation}
-2\bm{\Omega}\bm{\Delta}\left(\omega\right)=\bm{D}_{C}^{-1}\left(\omega\right)-\bm{D}_{\text{sys}}^{-1}\left(\omega\right)\,,
\end{equation}
where $\bm{D}_{C}\left(\omega\right)=N_{k}^{-1}\sum_{\bm{k}}\bm{D}\left(\bm{k},\omega\right)$ is the cellular GF, 
and $N_{k}$ is the number of points sampled in the BZ. 
Details regarding the definition of the impurity problem, including $V_{\text{loc}}(\bm{u})$, are given in the SM.

The (classical) dynamics of the system coordinates are governed by a set of coupled generalized Langevin equations (GLEs),
\begin{equation}
\ddot{\bm{u}}\left(t\right) = -\nabla V_{\text{eff}}\left( \bm{u} \right) - \int_{0}^{t}ds\bm{\gamma}\left(t-s\right)\dot{\bm{u}}\left(s\right)+\bm{\xi}\left(t\right)\,.\label{eq:gle}
\end{equation}
Here, $\bm{\gamma}(t)$ is a matrix of friction kernels that describes the dissipative effect of the bath on the system dynamics and is related to the hybridization, 
%
%
\begin{equation}
\bm{\gamma}\left(t\right)=-2\sqrt{\frac{2}{\pi}}\int_{0}^{\infty}d\omega\cos\left(\omega t\right)\frac{2\bm{\Omega}\Im\bm{\Delta}\left(\omega\right)}{\omega}\,,
\end{equation}
where $\Im(\cdot)$ denotes the imaginary part.
The effective potential, $V_\mathrm{eff}(\bm{u})$, includes the bath-induced renormalization of the local potential, 
and $\bm{\xi}(t)$ is a vector of random forces that satisfies the fluctuation-dissipation relation, $\left\langle \bm{\xi}\left(t\right)\bm{\xi}^T\left(s\right)\right\rangle=k_{B}T\bm{\gamma}\left(t-s\right)$.

As detailed in the SM, with thermal sampling of the initial conditions, the above GLEs are solved numerically~\cite{ceriotti_colored-noise_2010, ceriotti_-thermostat_2010} to obtain dynamics of the system coupled to the bath and to compute the anharmonic impurity GF, $\bm{D}_\mathrm{imp}(t) = (k_BT)^{-1} \theta(t) \langle \dot{\bm{u}}(t) \bm{u}^T(0)\rangle$.
%
%
%
%
From this, the self-energy is obtained as
\begin{equation}
\bm{\pi}\left(\omega\right)=\frac{1}{2}\bm{\Omega}^{-1}\left[\bm{D}_{\text{imp},0}^{-1}\left(\omega\right)-\bm{D}_{\text{imp}}^{-1}\left(\omega\right)\right]\,,\label{eq:seflfEnergy}
\end{equation}
where $\bm{D}_{\text{imp},0}(\omega)$ is the harmonic impurity GF.
%
This local self-energy is used to calculate the lattice GF [i.e., Eq.~(\ref{eq:latticeGF}) with the impurity $\bm{\pi}(\omega)$ in place of $\bm{\pi}(\bm{k},\omega)$], leading to an iterative procedure that converges once the self-consistency condition, $\bm{D}_{C}\left(\omega\right)=\bm{D}_{\text{imp}}\left(\omega\right)$, has been reached. For the systems studied here, we find that self-consistency is achieved in one iteration, as shown in the SM.

Due to the sampling of initial conditions, the calculated impurity GFs have statistical noise, 
leading to issues of non-causality and negative spectral functions (see SM). Furthermore, converging the numerical Fourier transform requires the propagation of long trajectories. 
Therefore, instead of numerically Fourier transforming $\bm{D}_{\text{imp}}(t)$, we fit its elements to the functional form of the GF of a damped harmonic oscillator and perform the Fourier transform analytically. This fitting technique, which is further described in the SM, circumvents the need to run many long trajectories---making our approach significantly more efficient while retaining excellent frequency resolution---and ensures a causal self-energy through simple constraints on the fitting parameters.

We use VDMFT to calculate the anharmonic GF and spectral function, $A\left(\bm{k},\omega\right)=-{\pi}^{-1}\text{Tr}\left[\Im\bm{D}\left(\bm{k},\omega\right)\right]$, of GG, BaGG, and SrGG at 300~K, which are illustrated in Fig.~\ref{fig:spectralFunction}. While the empty GG clathrate is largely harmonic, the filled clathrates feature anharmonicity that causes peaks to shift to energies higher than those predicted by the harmonic dispersion relation and causes them to broaden due to phonon scattering. Interestingly, anharmonicity affects modes differently; while the cage-dominant acoustic modes remain relatively unchanged from their harmonic dispersion, modes with appreciable guest character show large effects. In particular, the flat Ba(2) rattling modes shift from 36\,cm$^{-1}$ to 39\,cm$^{-1}$, and the Sr(2) rattling modes, which have lower harmonic frequencies and stronger quartic anharmonicity, shift from 21\,cm$^{-1}$ to 33\,cm$^{-1}$ and acquire a linewidth of 8\,cm$^{-1}$, which corresponds to a short lifetime of 0.67~ps. The hardening of these rattling modes also affects the avoided crossing with the cage acoustic modes.
In the SM, we present the temperature dependence of the VDMFT spectral function of SrGG from 50~K to 600~K, showing that the Sr(2) rattling mode significantly broadens and hardens with increasing temperature. 

\begin{figure}[t]
\includegraphics[width=3.375in]{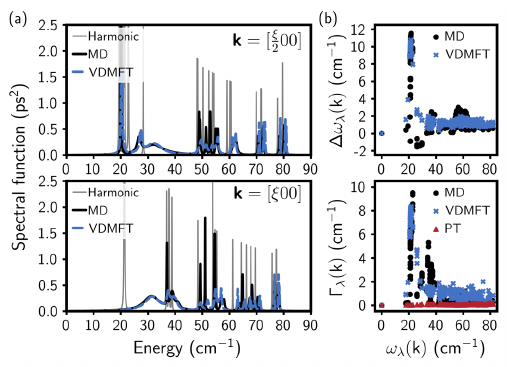}
\caption{(a) The spectral function of SrGG at 300~K at $\bm{k}=[\frac{\xi}{2} 0 0]$ and $\bm{k}=[\xi 0 0]$, where $\xi=\pi/a$, calculated using MD and VDMFT. (b) Anharmonic frequency shifts (top) and linewidths (bottom) of SrGG at 300~K obtained from the self-energy calculated using MD, VDMFT, and PT sampled on a $4\times 4 \times 4$ grid of the BZ. The frequency shift is calculated as $\Delta \omega_\lambda (\bm{k}) = \omega_{\text{eff},\lambda}(\bm{k}) - \omega_\lambda(\bm{k})$, where $\omega^2_{\text{eff},\lambda}(\bm{k}) = \omega^2_\lambda(\bm{k}) + 2\omega_\lambda(\bm{k}) \Re\bm{\pi}_{\lambda,\lambda} \big(\bm{k}, \omega_{\text{eff},\lambda}(\bm{k})\big)$ and was solved for iteratively. The linewidth is calculated as $\Gamma_\lambda (\bm{k}) = 2\omega_\lambda(\bm{k})\Im \bm{\pi}_{\lambda,\lambda} \big(\bm{k}, \omega_{\text{eff},\lambda} (\bm{k})\big) / \omega_{\text{eff}, \lambda} (\bm{k})$.
}
\label{fig:comparisonMD}
\end{figure}

To evaluate the accuracy of VDMFT, we compare to the exact spectral function computed using MD simulations of a large supercell with periodic boundary conditions (MD simulation details are given in the SM). Figures~\ref{fig:spectralFunction} and \ref{fig:comparisonMD} demonstrate that agreement between VDMFT and MD spectral functions is excellent for SrGG at 300~K, the most anharmonic system studied here. However, while MD offers limited resolution of the BZ due to the finite size of the simulated supercell, the VDMFT spectral function is accessible at all points in the BZ. Moreover, a VDMFT calculation is significantly more affordable than MD simulations: while using MD to compute the spectral function required the simulation of 8,192 atoms with periodic boundary conditions, using VDMFT required the simulation of only 8 atoms coupled to a bath.

Figure~\ref{fig:comparisonMD} shows the spectral function at specific points in the BZ as well as frequency shifts and linewidths calculated by both MD and VDMFT for SrGG at 300~K, again indicating the excellent accuracy of VDMFT in describing anharmonicity across the BZ. It is worth noting that VDMFT is accurate for various degrees of anharmonicity, capturing both the strong anharmonicity of the rattling modes as well as the nearly negligible anharmonicity of the higher-energy optical modes. Comparisons between VDMFT and MD for GG and BaGG are shown in the SM. Additionally, Fig.~\ref{fig:comparisonMD}b shows phonon linewidths computed using lowest-order PT of three-phonon scattering processes ~\cite{tadano2014anharmonic} (details in SM). As shown in the SM, PT predicts accurate linewidths for the quasi-harmonic GG, but it fails to capture any broadening for BaGG or SrGG, indicating that cage-guest anharmonicity in the filled clathrates is dominated by four- and higher-phonon scattering processes that cannot be described by PT but that are inherently included in VDMFT.

\begin{figure}[t]
\includegraphics[width=3.375in]{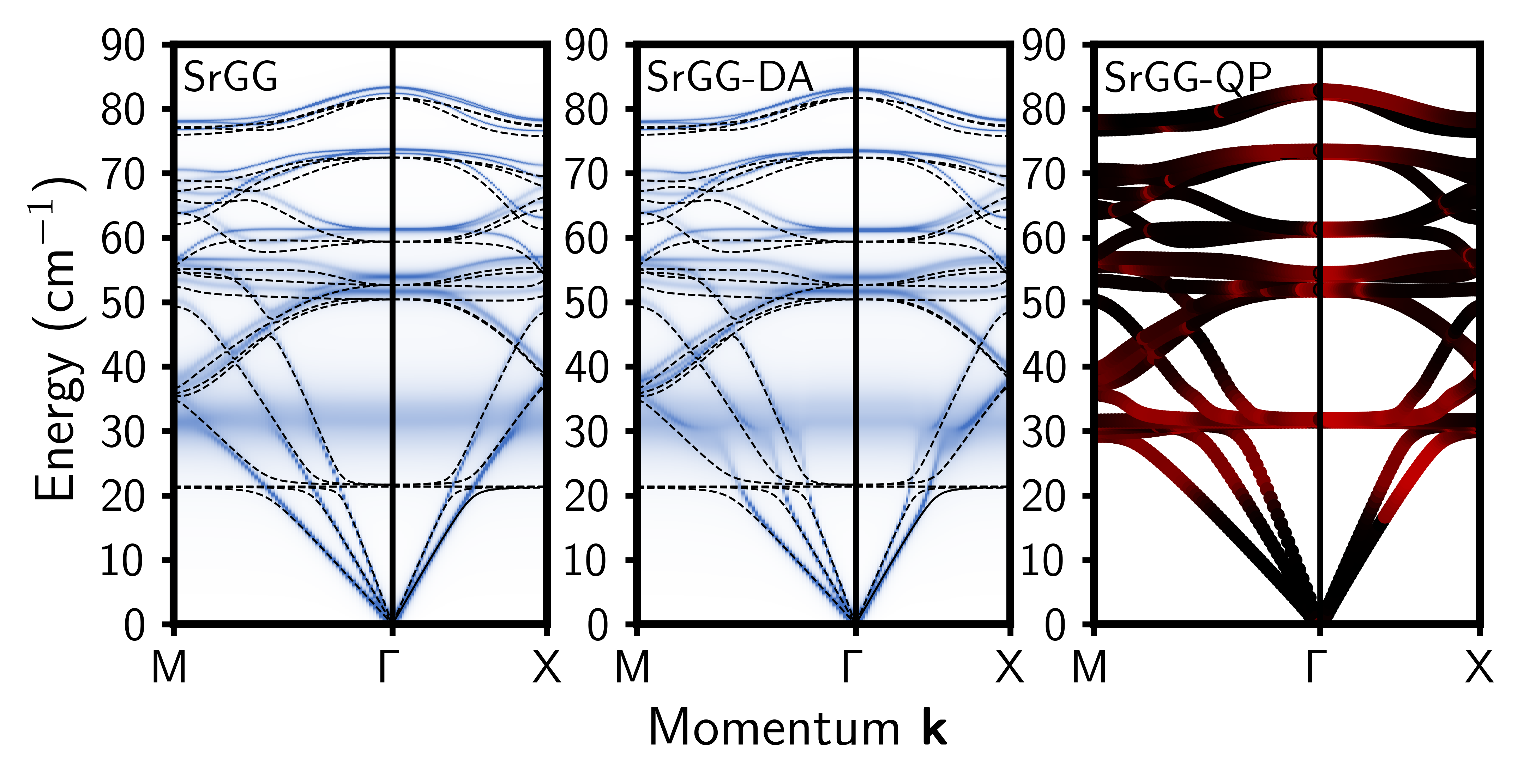}
\caption{The spectral functions of SrGG at 300~K calculated using the full self-energy (left), the diagonal approximation (DA, center), and the QP approximation to the full self-energy (right), where effective modes are colored according to the inverse participation ratio. Points colored in black indicate contribution from only a single harmonic mode while points colored in red indicate mixing of several harmonic modes.}
\label{fig:diagApprox}
\end{figure}

Next, we use our VDMFT results to better understand how anharmonicity impacts the phonon QP picture for SrGG at 300~K. To do this, we consider the diagonal approximation, which neglects non-diagonal elements of the self-energy in the phonon basis. 
Figure~\ref{fig:diagApprox} shows that the SrGG spectral function computed within the diagonal approximation deviates remarkably from the full spectral function, especially where the acoustic and rattling modes intersect. These results suggest that, in addition to shifting frequencies and imparting lifetimes, anharmonicity mixes the original phonon modes, i.e., those defined by the harmonic dynamical matrix. 
To quantify this anharmonic mode mixing, we calculate a static and Hermitian approximation to the VDMFT self-energy, as done for example in QP self-consistent GW~\cite{QPSGW_prl2006}, and use it to determine the improved phonon modes for SrGG at 300~K. The right panel of Fig.~\ref{fig:diagApprox} shows the band structure of these effective modes, which is in much better agreement with the peak positions of the fully anharmonic spectral function. In particular, the rattling mode is correctly shifted up by about 10~cm$^{-1}$. Through analysis of the inverse participation ratio of the QP bands, we find that cage acoustic and guest rattling modes between 25--35\,cm$^{-1}$ show significant mode mixing, as do modes near the BZ center that have both guest and cage character (Fig.~\ref{fig:model}). 
Some amount of mode-mixing can also be captured by static mean-field theory, which defines a more accurate QP basis that treats both local and nonlocal anharmonicity. Such an approach can be straightforwardly combined with VDMFT~\cite{shih_anharmonic_2022}, which would also increase its applicability to materials with longer range anharmonicity.


In conclusion, we have developed the methodological extensions of VDMFT for real materials, which we used to study the anharmonic lattice dynamics of clathrate solids. By comparison with exact MD simulations, we conclude that VDMFT is remarkably accurate and provides results of higher resolution at significantly lower cost. By comparison with conventional PT calculations, we find that anharmonicity in type-I clathrates is dominated by four-phonon and higher-order scattering processes, indicating that nonperturbative effects matter. This result is applicable to a wide range of symmetric host-framework structures, including perovskites, Heusler and half-Heusler compounds, and skutterudites, suggesting the importance of nonperturbative techniques in the accurate description of anharmonicity and related material properties, such as thermal conductivities.

The VDMFT approach introduced here is completely general and can be applied to any material, although it is best suited for those with strong, local anharmonicity. 
While this work uses a coarse-grained classical force field, VDMFT can be straightforwardly performed with all-atom force fields or \textit{ab initio} electronic structure theory, where the computational savings will be even more significant.

From our application to clathrate solids, we confirm that the introduction of guest atoms within the lattice framework leads to significant anharmonic effects, such as the hardening and broadening of phonon modes, that cannot be described by PT. 
Additionally, we find that anharmonicity changes the character of the phonon QPs via significant mixing between the X(2) rattling modes and cage acoustic modes. 
The impact of the strongly anharmonic rattling modes on the acoustic modes is known to have implications for thermal conductivities~\cite{christensen_avoided_2008,tadano_quartic_2018,ikeda_kondo-like_2019,godse_anharmonic_2022}, which can be computed using the VDMFT GF and will be the subject of future work. 

\textit{Acknowledgements.} We thank Petra Shih for helpful discussions.
This work was supported by the U.S. Department of Energy, Office of Science, Basic Energy Sciences, under Award
No. DE-SC0023002. We acknowledge computing resources
from Columbia University’s Shared Research Computing Facility project, which is supported by NIH Research Facility
Improvement Grant 1G20RR030893-01, and associated funds
from the New York State Empire State Development, Division
of Science Technology and Innovation (NYSTAR) Contract
C090171, both awarded April 15, 2010.

%

\end{document}


\title{Supplemental Material: Nonperturbative Simulation of Anharmonic Rattler Dynamics in Type-I Clathrates with Vibrational Dynamical Mean-Field Theory}

\author{Dipti Jasrasaria}
\email{dj2667@columbia.edu}
\address{Department of Chemistry, Columbia University, New York, New York 10027, USA}

\author{Timothy C. Berkelbach}
\email{t.berkelbach@columbia.edu}
\affiliation{Department of Chemistry, Columbia University, New York, New York 10027, USA}

\date{\today}

\maketitle

\section{Clathrate model and parameters}

Our clathrate model consists of large ``cage" atoms that form a face-centered cubic (FCC) lattice with a lattice constant of $a=10.95$\,\AA~and with four sites in each unit cell, whose coordinates are collected in Table~\ref{tab:fcc}. The cage atoms have a mass of $m_\text{cage}=1647.924$\,amu and interact with a Lennard-Jones (LJ) potential, which is defined by the parameters $\epsilon=2.7876$\,eV and $\sigma=7.0345$\,\AA~and a cutoff of $r_c=14.448$\,\AA.

\begin{center}
\begin{table}[h!]
    \caption{Fractional coordinates for the four sites in the FCC lattice unit cell. The lattice constant is $a=10.95$\,\AA.}
    \centering
    \begin{tabular}{cccc}
    \hline
       atom  & $x$ & $y$ & $z$ \\
    \hline
        $0$ & $0$ & $0$ & $0$ \\
        $1$ & $a/2$ & $a/2$ & $0$ \\
        $2$ & $a/2$ & $0$ & $a/2$ \\
        $3$ & $0$ & $a/2$ & $a/2$ \\
    \hline
    \end{tabular}
    \label{tab:fcc}
\end{table}
\par\end{center}

Additionally, at each lattice site is a smaller ``guest" atom that interacts with the cage atom at that site through an anharmonic, quartic potential. Note that guest-guest interactions are mediated through the cage framework. The Ba guest atoms have mass $m_\text{Ba}=137.327$\,amu, and the Sr guest atoms have mass $m_\text{Sr}=87.62$\,amu. The guest atoms at the cube vertices represent the guest atoms in dodecahedral cages and interact with the cage via an isotropic potential. The guest atoms at the cube faces represent the guest atoms in the tetrakaidecahedral cages and interact with the cage via a softer anisotropic potential. While the cage-guest interactions are anistropic, they are oriented in such a way that the overall crystal is isotropic, as shown schematically in Fig.~1b of the main text. We note that dynamical stability of identical LJ cage atoms requires the FCC structure, which only approximately captures the structure of the fully atomistic cages in type-I clathrates. In each unit cell, these materials have six tetrakaidecahedral cages and two dodecahedral cages. In our model, we have fused the six tetrakaidecahedral cages in a pairwise manner into the three cage atoms on the faces of the unit cell (analogously keeping only one of the two guest atoms), and we have kept only one of the two dodecahedral cages, on the corners of the unit cell, effectively losing the dodecahedral cage (and its guest) at the center of the unit cell. However, based on the good agreement between our harmonic phonon dispersions and the \textit{ab initio} ones from Refs.~\citep{tadano_impact_2015,tadano_quartic_2018,godse_anharmonic_2022}, we do not expect these differences to qualitatively impact our conclusions.

The quartic anharmonicity parameters were chosen to reproduce the behavior of \textit{ab initio} guest-cage potential energy surfaces reported in Ref.~\onlinecite{dong_chemical_2000} for guest atoms as a function of their displacement from the center of 20- and 24-atom cages. Because Ref.~\onlinecite{dong_chemical_2000} only reported results for highly symmetric I$_{2}$II$_{6}$Ga$_{14}$Ge$_{30}$, where I and II belong to the first and second groups of the periodic table, and the 20-atom cages only host group-I atoms, we adopted the following fitting procedure. Once the harmonic parameters were fixed through fitting of the harmonic dispersion relation, the $g_{\alpha,i}$ values were chosen to reproduce ratios of $K_{\alpha,i} / g_{\alpha,i}$ using the data presented in Ref.~\onlinecite{dong_chemical_2000} for the same or similar guest atoms.
The parameters defining the quartic potential for the cage-guest interactions at each of the four FCC sites are given in Table~\ref{tab:Vq}.


\begin{center}
\begin{table}[h!]
    \caption{Parameters defining the quartic cage-guest potential given by Eq.~(3) of the main text. The units for $K$ are eV$\cdot$\AA$^{-2}$, and the units for $g$ are eV$\cdot$\AA$^{-4}$}
    \centering
    \begin{tabular}{ccccccc}
    \hline
        atom & $K_x$ & $K_y$ & $K_z$ & $g_x$ & $g_y$ & $g_z$ \\
    \hline
        Ba$_0$ & 2.4820 & 2.4820 & 2.4820 & 0.6795 & 0.6795 & 0.6795 \\
        Ba$_1$ & 1.6563 & 0.6763 & 0.6763 & 0.8198 & 0.3550 & 0.3550 \\
        Ba$_2$ & 0.6763 & 0.6763 & 1.6563 & 0.3550 & 0.3550 & 0.8198 \\
        Ba$_3$ & 0.6763 & 1.6563 & 0.6763 & 0.3550 & 0.8198 & 0.3550 \\
    \hline
        Sr$_0$ & 1.6708 & 1.6708 & 1.6708 & 0.2293 & 0.2293 & 0.2293 \\
        Sr$_1$ & 0.9873 & 0.1483 & 0.1483 & 0.4555 & 0.2438 & 0.2438 \\
        Sr$_2$ & 0.1483 & 0.1483 & 0.9873 & 0.2438 & 0.2438 & 0.4555 \\
        Sr$_3$ & 0.1483 & 0.9873 & 0.1483 & 0.2438 & 0.4555 & 0.2438 \\
    \hline
    \end{tabular}
    \label{tab:Vq}
\end{table}
\par\end{center}

\section{Defining the impurity problem and local potential}

The unit cell that constitutes the system in the impurity problem has a local potential, $V_{\text{loc}}(\bm{u})$, which includes all harmonic and anharmonic interactions within the unit cell but only includes the local harmonic part of anharmonic interactions that cross cell boundaries \cite{shih_anharmonic_2022}. This definition of the local potential ensures that the self-energy obeys continuous translational symmetry because the local harmonic parts of the anharmonic interactions that cross cell boundaries, which break continuous translational symmetry in the anharmonic impurity GF, are canceled out exactly by the harmonic impurity GF, as seen in Eq. (11) of the main text.

As all cage-guest interactions are purely local, the local potential can be written as
\begin{align}
    V_\text{loc}(\bm{u}) = &\frac{1}{2}\sum_{\alpha,\beta}^{\text{cage}} V_{\text{LJ}} \big( | \bm{r}_\alpha - \bm{r}_\beta | \big) + \sum_{\alpha}^\text{cage} \sum_{\beta}^{\text{guest}}{}^{'} V_{\text{q}}^{\alpha \beta} (\bm{r}_\alpha - \bm{r}_\beta) \nonumber \\
    &-\frac{1}{2}\sum_{\alpha,\beta}^{\text{cage}}\sum_i \frac{\partial V_{\text{LJ}}}{\partial (u_{\alpha i} - u_{\beta i})}(u_{\alpha i} - u_{\beta i}) \nonumber \\
    &+ \frac{1}{2}\sum_{\alpha}^{\text{cage}} \sum_{\gamma}^{\text{cage}}{}^{''} \sum_{i,j} \frac{\partial^2 V_{\text{LJ}}}{\partial (u_{\alpha i} - u_{\gamma i}) \partial (u_{\alpha j} - u_{\gamma j})} u_{\alpha i} u_{\alpha j}\,,
    \label{eqn:Vloc}
\end{align}
where the primed summation indicates that only cages and guests on the same lattice site interact, and the double primed summation is over periodic images of atom $\alpha$ that are \textit{not} in the unit cell. Here, $\alpha,\beta$ are indices over atoms in the unit cell, and $i,j$ are indices over Cartesian directions. The instantaneous and equilibrium positions of atom $\alpha$ are given by $\bm{r}_\alpha$ and $\bm{R}_\alpha$, respectively, and the atomic displacement of atom $\alpha$ in direction $i$ is given by $u_{\alpha i} = r_{\alpha i} - R_{\alpha i}$. All derivatives are evaluated at the equilibrium configuration. The second line of Eq.~(\ref{eqn:Vloc}) subtracts linear terms of the cage-cage LJ interactions within the cell to ensure that the force on each atom in the system is zero at the equilibrium configuration. The third line describes the local harmonic part of the LJ interaction that each cage atom $\alpha$ has with cage atoms $\gamma$ that are outside of the cell, which depends on the curvature of the LJ potential at the equilibrium distance between atoms $\alpha$ and $\gamma$.

\section{Solving the impurity problem}

As described in the main text, the dynamics of the system coupled to the bath are governed by a set of generalized Langevin equations (GLEs)
with a local effective potential that includes the bath-induced renormalization of the local potential,
\begin{equation}
V_{\text{eff}}\left( \bm{u} \right)  =V_{\text{loc}}\left( \bm{u} \right)-\frac{1}{2\sqrt{2\pi}}\bm{u}^T\bm{\gamma}\left(t=0\right)\bm{u}\,.
\end{equation}

To solve the impurity dynamics, we write the GLEs given by Eq.~(9) of the main text in the basis of normal modes of the isolated system, $u_\lambda =\sum_{\alpha i} c_{\alpha i,\lambda} \sqrt{m_\alpha} u_{\alpha i}$, where $c_{\alpha i,\lambda}$ is the $\alpha i$ element of the $\lambda$ eigenvector of $\bm{\Omega}^2$. In this basis, non-diagonal elements of the friction kernel are small, and so we neglect them (i.e., $\gamma_{\lambda,\lambda'}\left(t\right)=\delta_{\lambda,\lambda'}\gamma_{\lambda,\lambda}\left(t\right)$), which simplifies the GLEs, but we emphasize that the normal modes are still coupled by local anharmonicity. Thus, the GLE for a given coordinate is given by
\begin{equation}
\ddot{u}_{\lambda}\left(t\right)  =-\frac{\partial V_{\text{eff}}\left(\bm{u} \right)}{\partial u_{\lambda}}-\int_{0}^{t}ds\gamma_{\lambda}\left(t-s\right)\dot{u}_{\lambda}\left(s\right)+\xi_{\lambda}\left(t\right)\label{eq:gle}\,,
\end{equation}
where $\gamma_\lambda$ is the friction kernel that describes the dissipative effect of the bath on coordinate $u_\lambda$:
\begin{equation}
\gamma_\lambda = -2\sqrt{\frac{2}{\pi}} \int_0^\infty d\omega \cos(\omega t) \frac{[2\bm{\Omega}\Im\bm{\Delta}]_{\lambda,\lambda} \left(\omega\right)}{\omega}\,,
\end{equation}
and $\xi_\lambda$ is a random force that satisfies the fluctuation-dissipation relation, $\left\langle \xi_{\lambda}\left(t\right)\xi_{\lambda}\left(s\right)\right\rangle =k_{B}T\gamma_{\lambda}\left(t-s\right)$.

We solve these GLEs numerically by rewriting each as a Markovian stochastic differential equation in an extended phase space through the addition of $n$ fictitious degrees of freedom, $\bm{s}_{\lambda}$~\cite{ceriotti_colored-noise_2010, ceriotti_-thermostat_2010}:
\begin{equation}
\left(\begin{array}{c}
\ddot{u}_{\lambda}\\
\dot{\bm{s}}_{\lambda}
\end{array}\right)=\left(\begin{array}{c}
-\frac{\partial V_{\text{eff}}\left(\left\{ u_{\lambda}\right\} \right)}{\partial u_{\lambda}}\\
0
\end{array}\right)-\bm{A}_{\lambda}\left(\begin{array}{c}
\dot{u}_{\lambda}\\
\bm{s}_{\lambda}
\end{array}\right)+\bm{B}_{\lambda}\left(\bm{\zeta}_{\lambda}\right)\,,\label{eq:gle_markovian}
\end{equation}
where $\bm{A}_{\lambda}$ is a matrix determined from a functional fitting of $\gamma_{\lambda}\left(\omega\right)$, $\bm{B}_{\lambda}\bm{B}_{\lambda}^{T}=k_{B}T$$\left(\bm{A}_{\lambda}+\bm{A}_{\lambda}^{T}\right)$, and $\bm{\zeta}_\lambda$ is a vector of $n+1$ uncorrelated Gaussian random numbers. To determine the elements of $\bm{A}_{\lambda}$, we fit the friction kernel to a functional form \cite{ceriotti_-thermostat_2010},
\begin{equation}
\gamma_{\lambda}\left(\omega\right)=\frac{1}{2\pi}\sum_{i=1}^{m}\eta_{i}\gamma_{i}\frac{\omega^{2}+\omega_{i}^{2}+\gamma_{i}^{2}}{\left[\left(\omega+\omega_{i}\right)^{2}+\gamma_{i}^{2}\right]\left[\left(\omega-\omega_{i}\right)^{2}+\gamma_{i}^{2}\right]}\,,
\end{equation}
and use the parameters $\{ \eta_i, \gamma_i, \omega_i \}$ to specify the elements of $\bm{A}_{\lambda}$:
\begin{equation}
\bm{A}_{\lambda}=\left(\begin{array}{cccccc}
0 & \sqrt{\eta_{1}/2\pi} & \sqrt{\eta_{1}/2\pi} & \cdots & \sqrt{\eta_{m}/2\pi} & \sqrt{\eta_{m}/2\pi}\\
-\sqrt{\eta_{1}/2\pi} & \gamma_{1} & \omega_{1} & \cdots & 0 & 0\\
-\sqrt{\eta_{1}/2\pi} & -\omega_{1} & \gamma_{1} & \cdots & 0 & 0\\
\vdots & \vdots & \vdots & \ddots\\
-\sqrt{\eta_{m}/2\pi} & 0 & 0 &  & \gamma_{m} & \omega_{m}\\
-\sqrt{\eta_{m}/2\pi} & 0 & 0 &  & -\omega_{m} & \gamma_{m}
\end{array}\right)\,.
\end{equation}
The elements of $\bm{B}_{\lambda}$ are determined by the fluctuation dissipation theorem and do not need to be set explicitly. For the systems studied here, the spectral densities can be fit using a sum of $m=6$ functions, resulting in the addition of $n=12$ fictitious degrees of freedom per normal mode.

We numerically propagate an ensemble of trajectories, where initial configurations are sampled from a Boltzmann distribution using Monte Carlo sampling, and initial momenta (for both real and fictitious degrees of freedom) are sampled from a Maxwell-Boltzmann distribution, and calculate the anharmonic impurity Green's function (GF):
\begin{equation}
    \big[\bm{D}_{\text{imp}}\big]_{\lambda,\lambda'} \left(t\right)=\frac{\theta\left(t\right)}{k_B T}\left\langle \dot{u}_{\lambda}\left(t\right)u_{\lambda'}\left(0\right)\right\rangle \,.
\end{equation}
The GFs are calculated by averaging over 500,000 trajectories of 20 ps each with a timestep of 0.01 ps. The harmonic impurity GF, $\bm{D}_{\text{imp},0}(t)$ can be computed in the same way by replacing $V_\text{loc}(\bm{u})$ with the harmonic version of Eq.~(\ref{eqn:Vloc}). 

\section{Fitting impurity Green's functions and enforcing causality}

\begin{figure}[h]
\includegraphics[width=5.0625in]{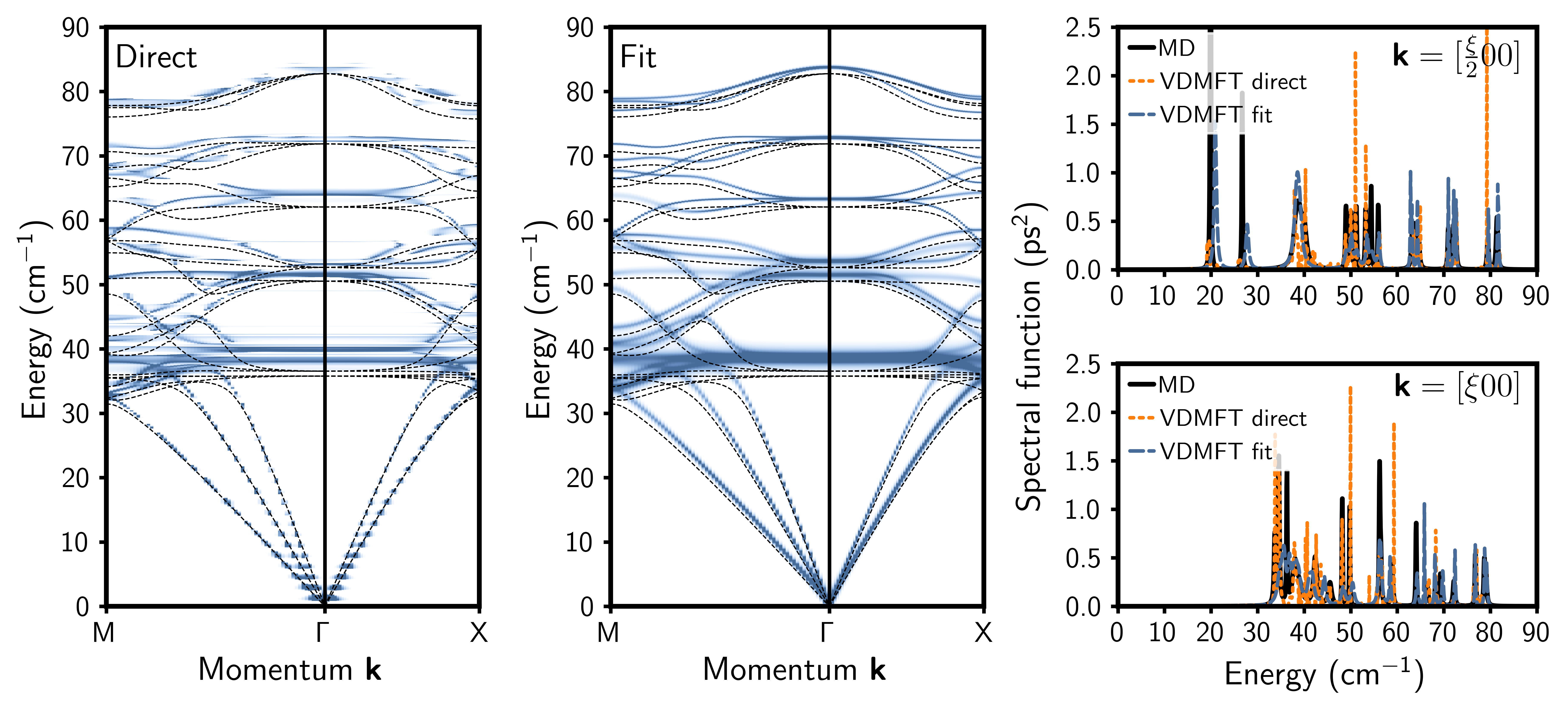}
\caption{The spectral function of BaGG at 300K using VDMFT. When the Fourier transforms of the impurity GFs are performed numerically, stochastic noise in the GFs is magnified and leads to causality issues (left). These issues can be addressed by fitting the time-domain GFs to a functional form and performing the Fourier transform analytically, resulting in a fully causal self-energy and positive spectral function (\textcolor{black}{center}). \textcolor{black}{The spectral function at specific cuts through the BZ at $\bm{k}=[\frac{\xi}{2} 0 0]$ (top right) and $\bm{k}=[\xi 0 0]$ (bottom right), where $\xi=\pi/a$, calculated using MD and VDMFT. The dotted orange lines show the direct method, where Fourier transforms are performed numerically, while the dashed blue lines show the fitting method, where the Fourier transforms are performed analytically.}}
\label{fig:causality}
\end{figure}

Due to the sampling of initial conditions, the calculated impurity GFs have statistical noise. This noise is magnified in the self-energy, which depends on the difference between two inverse GFs, leading to issues of non-causality and spectral functions with areas of negative spectral weight (shown in the left panel of Fig.~\ref{fig:causality}). Furthermore, converging the numerical Fourier transform requires the propagation of long trajectories. \textcolor{black}{The unconverged, numerically Fourier transformed spectral functions generally feature peaks at the correct frequencies, as illustrated in the right panel of Fig.~\ref{fig:causality}, but they are subject to significant noise such that the determination of linewidths and peak heights is not possible.}

To address these challenges, we fit each element of the impurity GFs to the functional form of the GF of a damped harmonic oscillator and perform the Fourier transform analytically. This fitting technique circumvents the need to run several long trajectories, making our approach more efficient, and ensures causality through simple constraints on the fitting parameters. However, more complicated functional forms may be necessary to capture strongly asymmetric peaks or satellite features.

\begin{figure}[h]
\includegraphics[width=5.0625in]{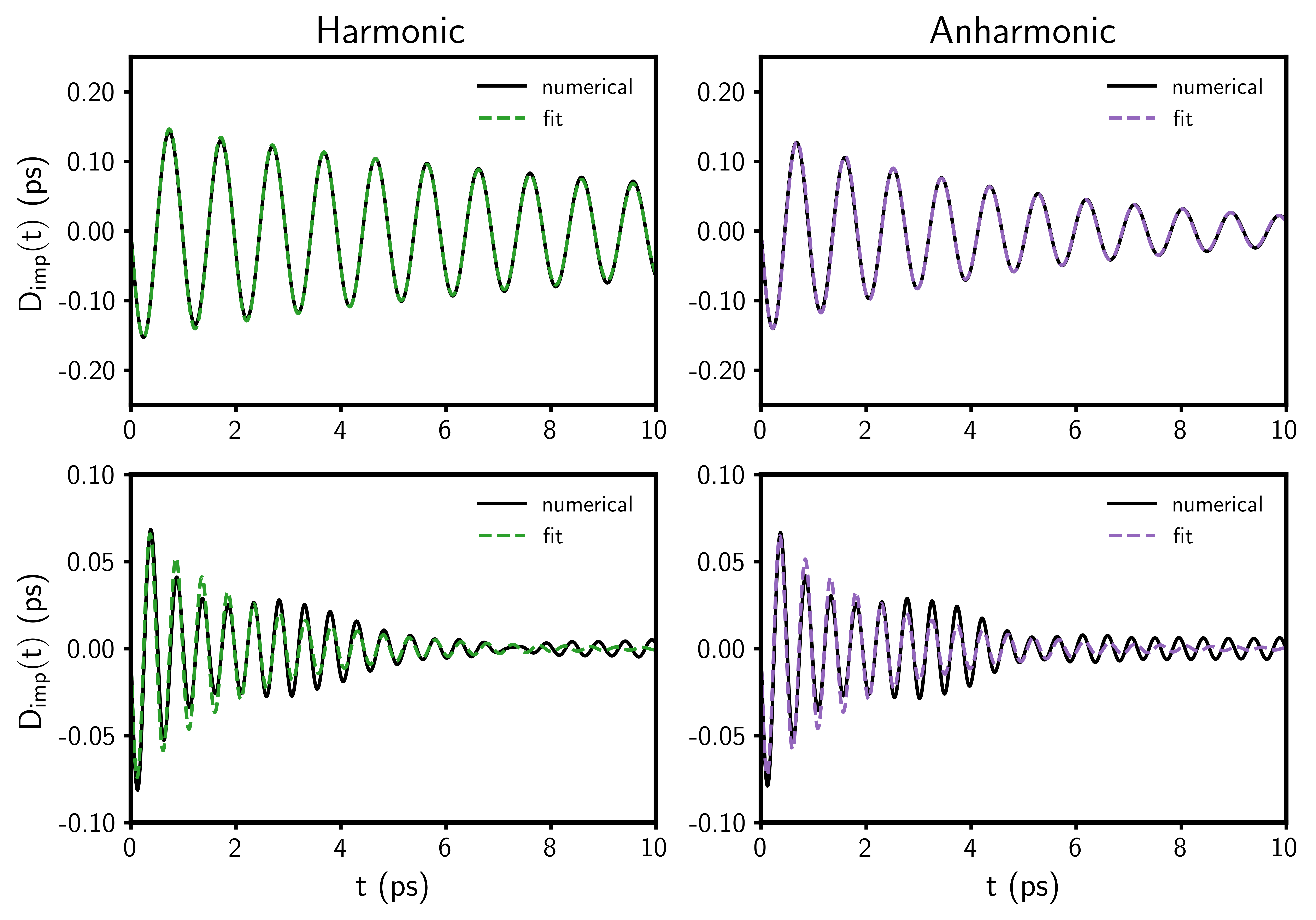}
\caption{The numerical and fit impurity GFs for two different coordinates in BaGG at 300K. Harmonic impurity GFs are shown on the left while anharmonic impurity GFs are shown on the right.}
\label{fig:fitting}
\end{figure}

Each element of both the harmonic and anharmonic impurity GFs is fit using two free parameters, $\omega$ and $\gamma$,
\begin{equation}
    \big[\bm{D}_{\text{imp}}\big]_{\lambda,\lambda'} (t) = - \frac{\theta(t)}{\Omega_{\lambda,\lambda'}} \exp(-\gamma_{\lambda,\lambda'}t/2)\sin(\Omega_{\lambda,\lambda'} t)\,,
    \label{eq:ChiFit}
\end{equation}
where $\Omega_{\lambda,\lambda'}^2 = \omega_{\lambda,\lambda'}^2 - \frac{1}{4}\gamma_{\lambda,\lambda'}^2$. Examples of GFs and their fits are given in Fig.~\ref{fig:fitting}.

While non-diagonal elements of $\bm{D}_{\text{imp},0}$ are strictly zero, non-diagonal elements of $\bm{D}_{\text{imp}}$ are generally non-zero because degrees of freedom are coupled to one another through the anharmonic potential. However, the magnitudes of these non-diagonal elements are much smaller than those of the diagonal elements for GG and BaGG, so we neglect them. Thus, the parameters used to fit the diagonal elements of the impurity GFs can be used to calculate the self-energy term directly:
\begin{equation}
\left[\bm{\Omega}\bm{\pi}\right]_{\lambda,\lambda}\left(\omega\right)=\left(\omega_{\lambda,\text{h}}^{2}-\omega_{\lambda,\text{anh}}^{2}\right)-i\omega\left(\gamma_{\lambda,\text{h}}-\gamma_{\lambda,\text{anh}}\right)\,.\label{eq:piFit}
\end{equation}
This form allows for the enforcement of causality of the self-energy through a simple constraint of the fitting parameters, $\gamma_{\lambda,\text{anh}}>\gamma_{\lambda,\text{h}}$, guaranteeing positive spectral functions, as illustrated in Fig.~\ref{fig:causality}.

For more anharmonic systems, such as SrGG, certain degrees of freedom are negligibly coupled to the bath but are strongly coupled to one other via the anharmonic potential, resulting in non-diagonal elements of $\bm{D}_{\text{imp}}$ that are of the same order of magnitude as the diagonal elements. In those cases, we fit the non-diagonal elements to the same form as the one given in Eq. (\ref{eq:ChiFit}) but with the addition of a parameter that multiplies the function.

With the inclusion of non-diagonal elements to $\bm{D}_{\text{imp}}$, there exists no straightforward expression for the self-energy directly from fitting parameters, so we calculate it using Eq. (11) of the main text. Furthermore, there is no constraint on fitting parameters to ensure causality, and small non-causal features may exist. To address this non-causality, we consider the spectral function of the anharmonic system,
\begin{equation}
    \bm{A}_{\text{sys}}\left(\omega\right)=-\frac{1}{\pi}\Im\bm{D}_{\text{sys}}\left(\omega\right)\,,    
\end{equation}
which should be positive semi-definite at all frequencies. At each frequency, we 
remove negative eigenvalues from an eigenvalue decomposition of $\bm{A}_{\text{sys}}\left(\omega\right)$.
Then, we obtain the real part of $\bm{D}_{\text{sys}}$ using Kramers-Kronig relations and extract the causal self-energy from $\bm{D}_{\text{sys}}^{-1}$.

\section{Calculating phonon Green's functions with molecular dynamics}

In the classical limit assumed here, the exact anharmonic dynamics of the lattice can be modeled using molecular dynamics (MD) simulations of a large supercell with periodic boundary conditions. From these simulations, we compute the anharmonic lattice GF:
\begin{equation}
\bm{D}_{\lambda,\lambda'}\left(\bm{k},t\right) = \frac{\theta(t)}{k_B T}\left\langle \dot{u}_{\lambda}\left(\bm{k},t\right)u_{\lambda'}\left(-\bm{k},0\right)\right\rangle\,.
\end{equation}
The GFs are calculated by averaging over 20,000 trajectories of 20 ps each with a timestep of 0.0025 ps. Initial configurations were sampled at intervals of 25 ps from an MD trajectory, where the temperature was controlled using a Langevin thermostat. MD simulations were performed using the LAMMPS code~\cite{Plimpton1995}. Again, we fit each element of the GF to the functional form given in Eq. (\ref{eq:ChiFit}) and analytically perform a Fourier transform to obtain the frequency-domain GF and then compute the spectral function. However, it is important to note that the Brillouin zone (BZ) sampling resolution is dictated by the size of the simulated supercell. Calculating the phonon GF from MD simulations with large supercells can be prohibitively expensive, severely limiting the resolution of the BZ sampling. To obtain the GF at $\bm{k}$-points along the $\Gamma-X$ path of the BZ we used a simulation supercell size of $64\times 4 \times 4$ unit cells, and to obtain the GF at $\bm{k}$-points along the $\Gamma-M$ path of the BZ we used a simulation supercell size of $32\times 32 \times 1$ unit cells.

\section{Calculating phonon linewidths using perturbation theory}

To calculate phonon linewidths using lowest-order perturbation theory (PT) on the three-phonon scattering, we first estimate harmonic and anharmonic interatomic force constants (IFCs). For each clathrate system, first we fit harmonic IFCs using the finite displacement approach~\cite{Parlinski1997} with a reference data set generated by displacing atoms from their equilibrium positions in a 3$\times$3$\times$3 supercell with displacement lengths of 0.01\,\AA, 0.02\,\AA, and 0.03\,\AA. Next, we simultaneously estimate anharmonic IFCs up to sixth order using the compressive sensing lattice dynamics method~\cite{Zhou2014} using a reference data set of 1000 configurations sampled from an equilibrium MD trajectory of a 3$\times$3$\times$3 supercell at 600~K. The temperature was controlled using a Langevin thermostat, and configurations were sampled at intervals of 2.5\,ps to ensure that they were not correlated. We consider all harmonic and anharmonic IFCs between cage atoms within the cutoff radius of the LJ potential ($r_c=14.448$\,\AA) and cage-guest pairs at the same lattice site.

\begin{figure}[h]
\includegraphics[width=5.0625in]{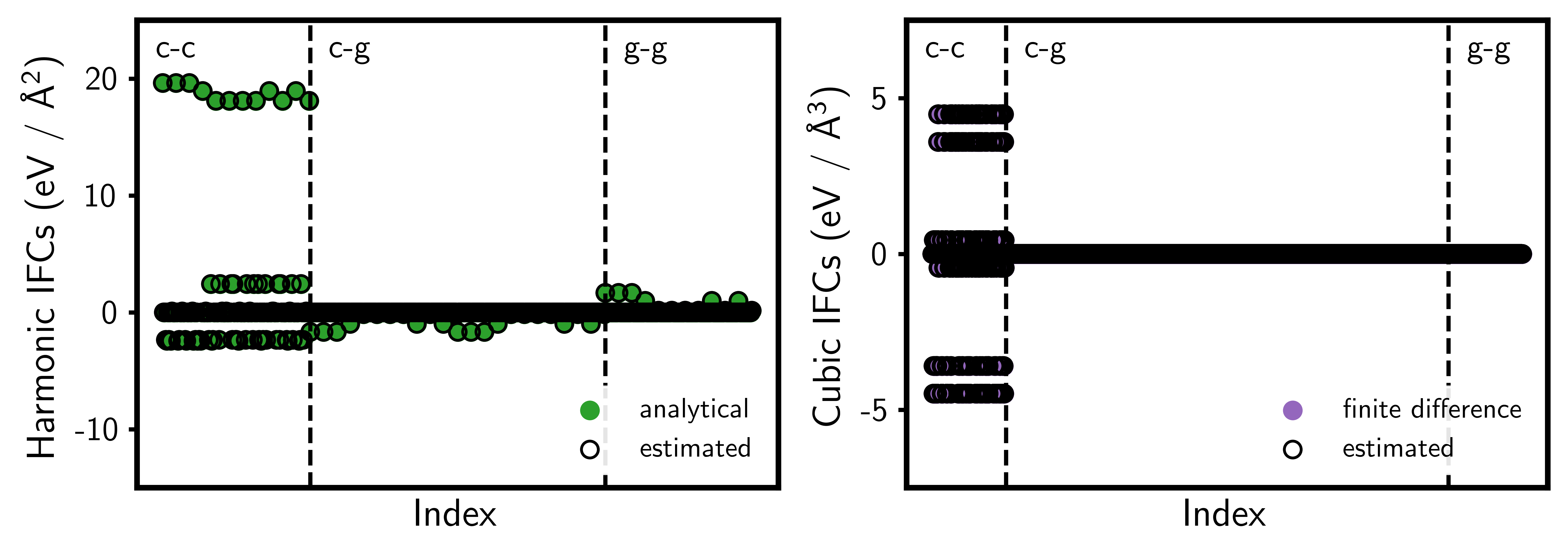}
\caption{The harmonic (left) and cubic (right) IFCs for SrGG. Cage-cage interactions are denoted by the label `c-c', cage-guest interactions by `c-g', and guest-guest by `g-g.'}
\label{fig:ifcs}
\end{figure}

To verify the accuracy of the estimated harmonic and third-order IFCs, which are used in the PT calculation, we compare them to the analytical harmonic IFCs and third-order IFCs calculated using finite differences of the analytical harmonic IFCs. We find excellent agreement, as shown for SrGG in Fig.~\ref{fig:ifcs}. 

Next, we use lowest-order PT to calculate the phonon linewidths as~\cite{Maradudin1962}:
\begin{align}
    \Gamma_\lambda(\bm{k},\omega) = &\frac{\pi}{2N} \sum_{\bm{k}',\bm{k}''} \sum_{\lambda',\lambda''} \big\vert V_{\lambda,\lambda',\lambda''}^{(3)} (-\bm{k},\bm{k}',\bm{k}'')\big\vert^2 \nonumber \\
    &\times\big[ \big(n_{\lambda'}(\bm{k}') + n_{\lambda''}(\bm{k}'') + 1\big)\delta\big(\omega-\omega_{\lambda'}(\bm{k}')-\omega_{\lambda''}(\bm{k}'')\big) \nonumber \\
    &-2\big(n_{\lambda'}(\bm{k}') - n_{\lambda''}(\bm{k}'')\big)\delta\big(\omega-\omega_{\lambda'}(\bm{k}')+\omega_{\lambda''}(\bm{k}'')\big)\big]\,,
\end{align}
where $\omega_\lambda(\bm{k})$ is the harmonic frequency of mode $\lambda$ at $\bm{k}$, $N$ is the number of points sampled in the BZ, and $n_\lambda(\bm{k})=[\exp(\hbar\omega_\lambda(\bm{k})/k_B T)-1]^{-1}$ is the Bose-Einstein distribution. The matrix element $V^{(3)}$ is the three-phonon interaction:
\begin{align}
    V_{\lambda,\lambda',\lambda''}^{(3)}(\bm{k},\bm{k}',\bm{k}'') = &\bigg(\frac{\hbar^3}{8N^2\omega_\lambda(\bm{k})\omega_{\lambda'}(\bm{k}')\omega_{\lambda''}(\bm{k}'')}\bigg)^{1/2} \nonumber \\
    & \times \sum_{\alpha,\beta,\gamma} \frac{1}{\sqrt{m_\alpha m_\beta m_\gamma}} \sum_{i,j,k} c_{\alpha i,\lambda}(\bm{k}) c_{\beta j,\lambda'}(\bm{k}') c_{\gamma k,\lambda''}(\bm{k}'') \nonumber \\
    & \times \sum_{\bm{l},\bm{m},\bm{n}} e^{i(\bm{k}\cdot\bm{R}_{\bm{l}\alpha} + \bm{k}'\cdot\bm{R}_{\bm{m}\beta} + \bm{k}''\cdot\bm{R}_{\bm{n}\gamma})}\frac{\partial^3 \bm{\mathcal{V}}}{\partial u_{\bm{l}\alpha i} \partial u_{\bm{m}\beta j} \partial u_{\bm{n}\gamma k}}\,,
\end{align}
where $\bm{l},\bm{m},\bm{n}$ are lattice translation vectors, $\alpha,\beta,\gamma$ are indices over atoms in the unit cell, and $i,j,k$ are indices over the Cartesian directions. $\bm{R}_{\bm{l}\alpha}$ is the equilibrium position of atom $\alpha$ in cell $\bm{l}$. The derivative of the lattice potential, $\bm{\mathcal{V}}$, with respect to atomic displacements is the third-order IFC, and $\bm{c}(\bm{k})$ are the eigenvectors of the dynamical matrix $\bm{\mathcal{D}}(\bm{k})$. 
To calculate the lifetimes, we use a 16$\times$16$\times$16 $\bm{k}$-point grid and evaluate the Dirac delta functions using the tetrahedron method. All IFC estimation and PT calculations were performed using the ALAMODE code package~\cite{tadano2014anharmonic}.

\begin{figure}[h]
\includegraphics[width=5.0625in]{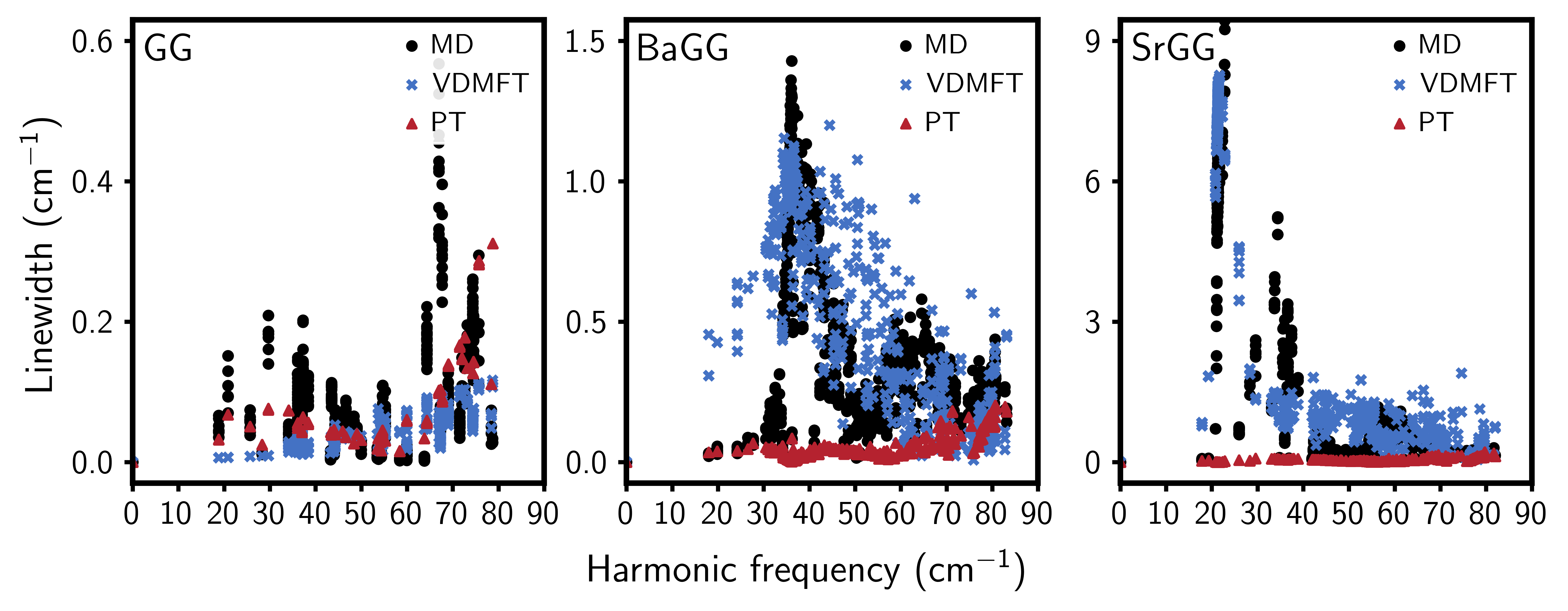}
\caption{Anharmonic linewidths of GG (left), BaGG (center), and SrGG (right) at 300 K obtained from the self-energy calculated using MD and VDMFT and calculated using PT, sampled on a 4$\times$4$\times$4 grid of the BZ.}
\label{fig:compareMD-PT}
\end{figure}

The linewidths calculated for GG, BaGG, and SrGG at 300~K sampled on a 4$\times$4$\times$4 grid of the BZ are illustrated in Fig.~\ref{fig:compareMD-PT} and compared to linewidths calculated using MD and VDMFT, as described in the main text. While PT accurately predicts the phonon linewidths for the quasi-harmonic GG, it fails to predict any significant broadening for the filled clathrates, BaGG and SrGG. Due to the cubic symmetry of our model clathrates and the anharmonic cage-guest interactions, all on-site third-order IFCs are zero and are unaffected by cage-guest interactions (Fig.~\ref{fig:ifcs}). Thus, the three-phonon interaction matrix elements of guest-dominant modes are extremely small, resulting in very small linewidths. This can be seen especially in the X(2) rattling modes with harmonic frequencies of 38\,cm$^{-1}$ for BaGG and 21\,cm$^{-1}$ for SrGG, which have broad linewidths that are accurately described by MD and VDMFT but that are incorrectly predicted to be very narrow by lowest-order PT.

\pagebreak
\section{Temperature dependence of VDMFT spectral functions}

\begin{figure*}[h!]
\includegraphics[width=6.75in]{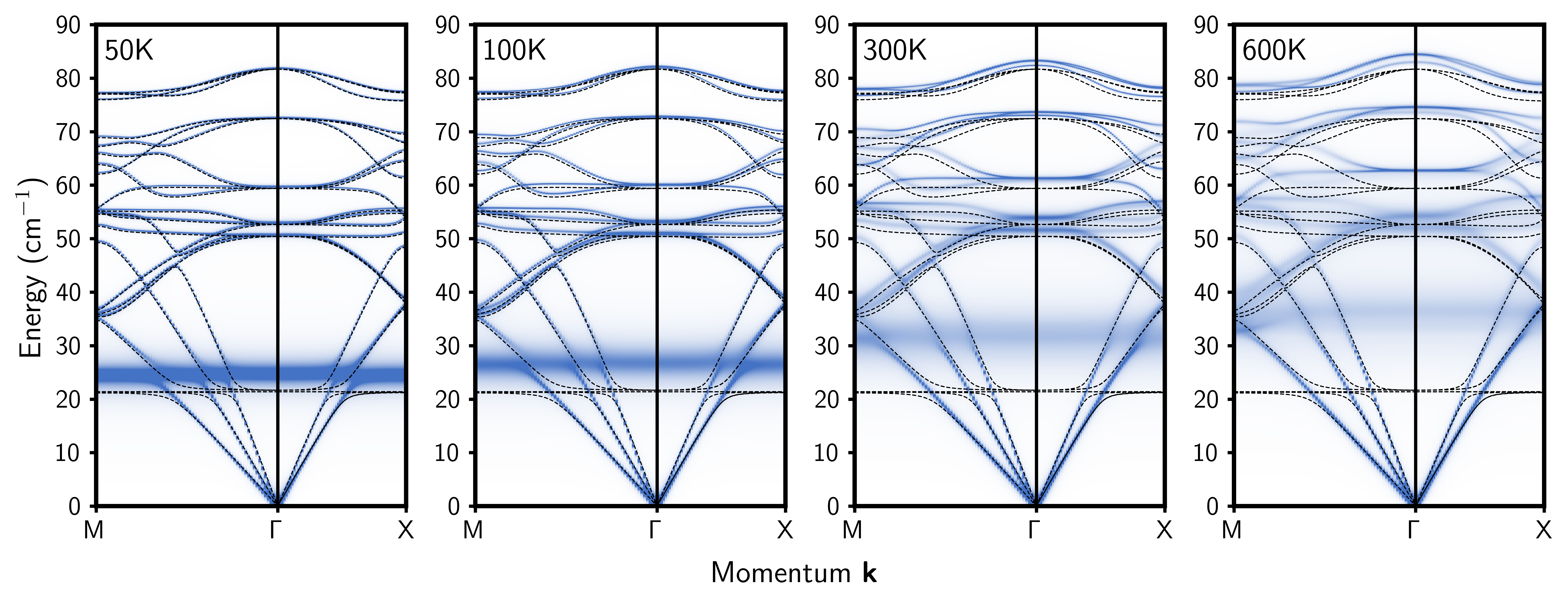}
\caption{Spectral functions of SrGG at 50~K, 100~K, 300~K, and 600~K calculated using VDMFT. Black dashed lines indicate the harmonic dispersion relation.}
\label{fig:spectralFunction}
\end{figure*}

\section{Comparing spectral functions calculated using VDMFT and MD}

\begin{figure}[hb]
\includegraphics[width=5.0625in]{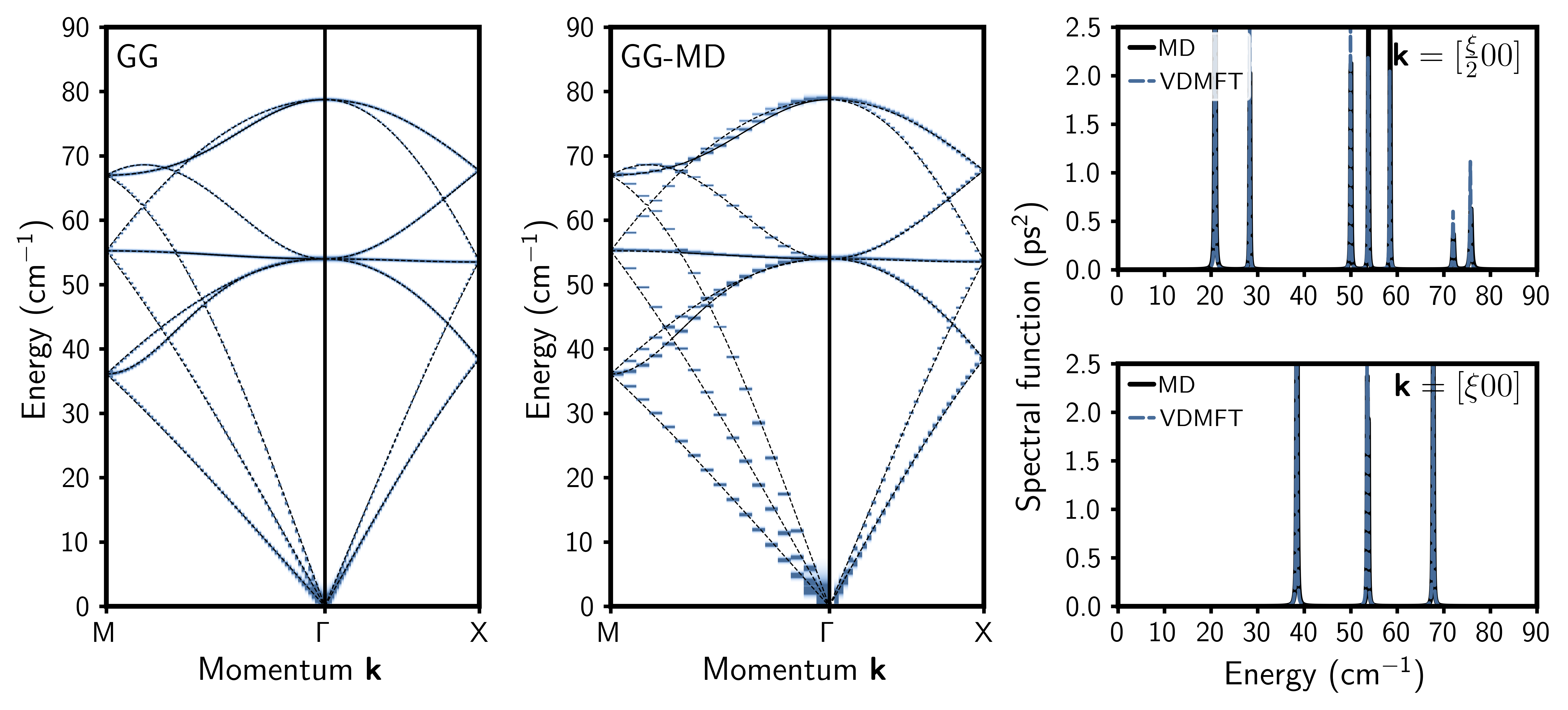}
\caption{The spectral function of GG at 300K calculated using VDMFT (left) and MD (center). The spectral function at specific cuts through the BZ at $\bm{k}=[\frac{\xi}{2} 0 0]$ (top right) and $\bm{k}=[\xi 0 0]$ (bottom right), where $\xi=\pi/a$, calculated using MD and VDMFT.}
\label{fig:compare}
\end{figure}

\begin{figure}[h]
\includegraphics[width=5.0625in]{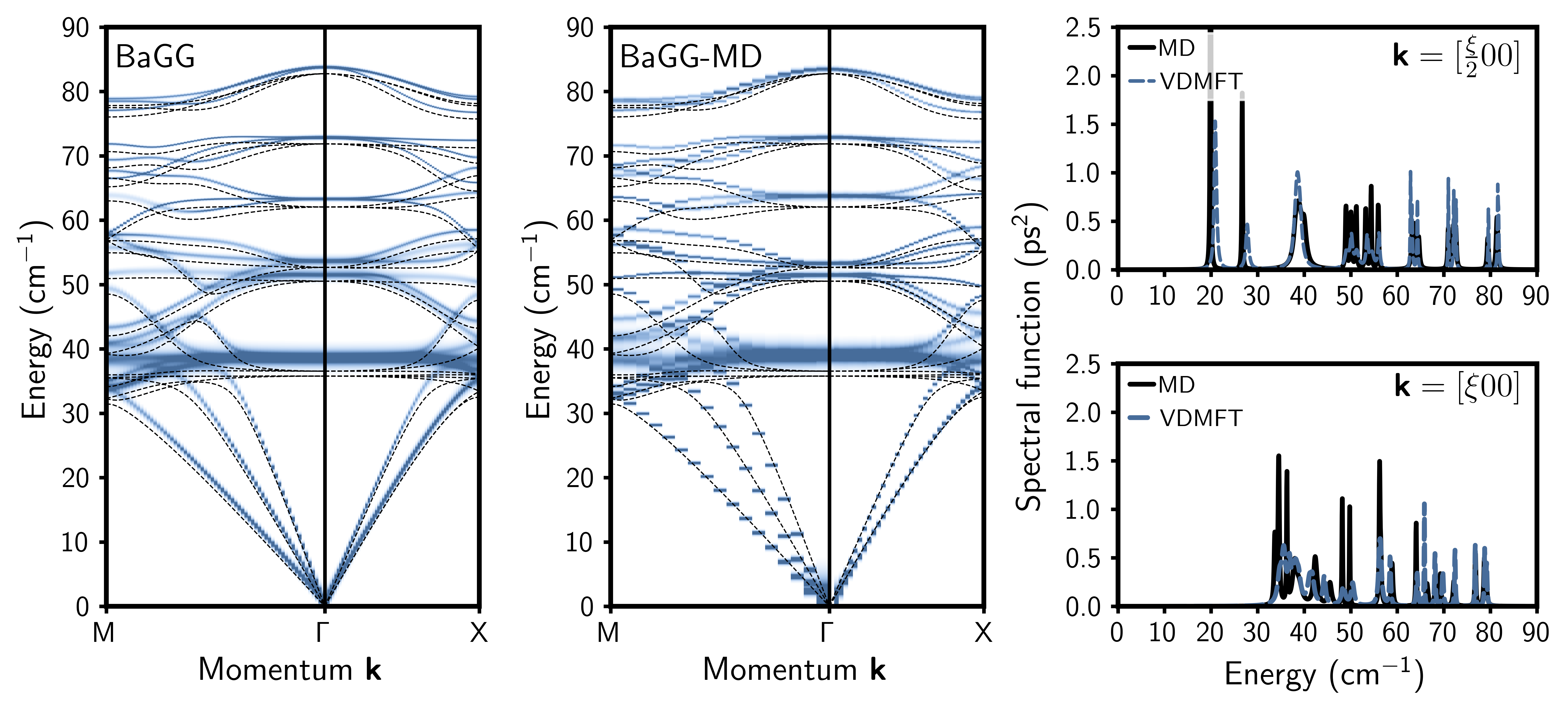}
\caption{The spectral function of BaGG at 300K calculated using VDMFT (left) and MD (center). The spectral function at specific cuts through the BZ at $\bm{k}=[\frac{\xi}{2} 0 0]$ (top right) and $\bm{k}=[\xi 0 0]$ (bottom right), where $\xi=\pi/a$, calculated using MD and VDMFT.}
\label{fig:compare}
\end{figure}

\pagebreak
\section{Assessing self-consistency of VDMFT}

\begin{figure}[h]
\includegraphics[width=5.0625in]{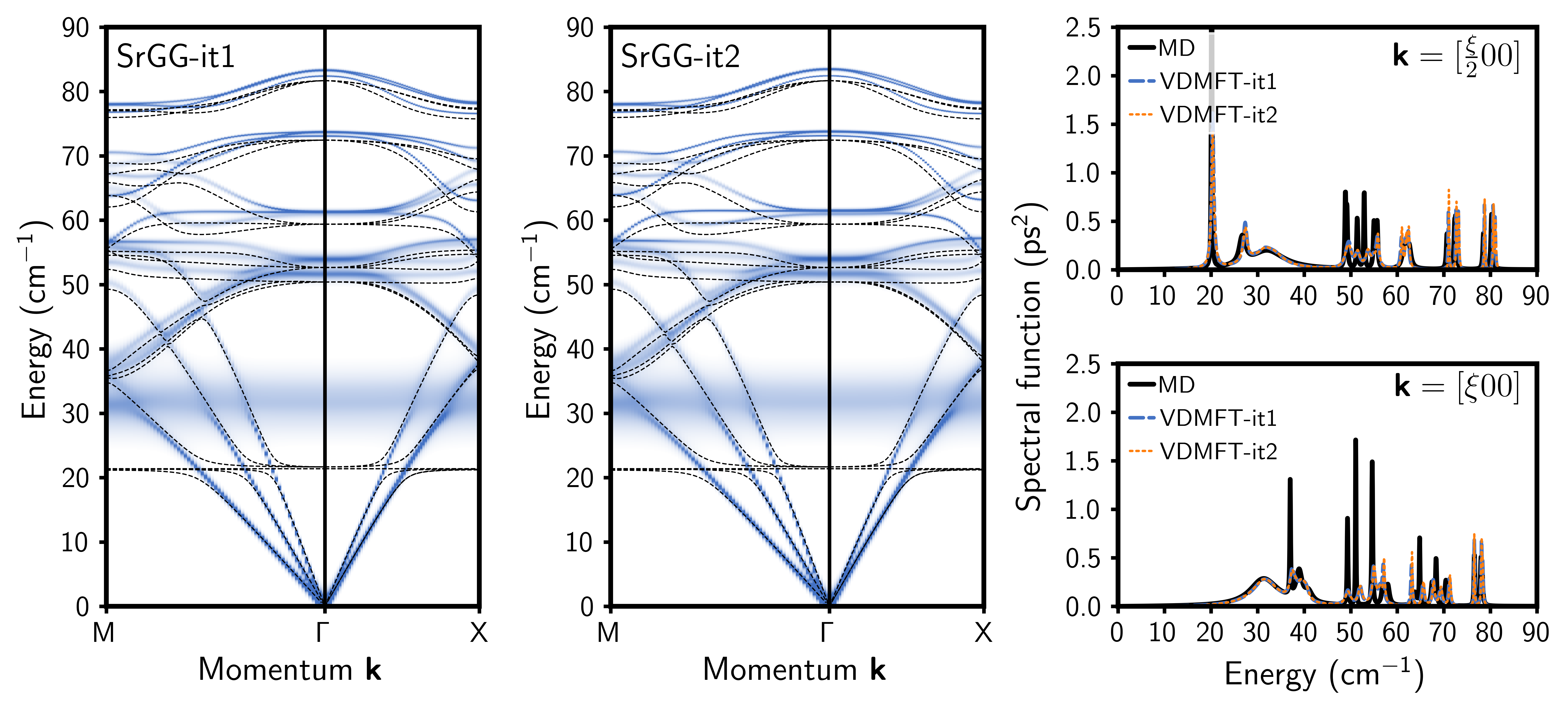}
\caption{The spectral function of SrGG at 300K calculated using VDMFT after the first iteration (left) and second iteration (center). The spectral function at specific cuts through the BZ at $\bm{k}=[\frac{\xi}{2} 0 0]$ (top right) and $\bm{k}=[\xi 0 0]$ (bottom right), where $\xi=\pi/a$, calculated using MD and VDMFT after the first iteration and second iteration.}
\label{fig:convergence}
\end{figure}

In Fig.~\ref{fig:convergence}, we show the VDMFT spectral functions of SrGG at 300~K after the first and second iterations, which are essentially identical, demonstrating that VDMFT has converged.

%